\documentclass[conference]{IEEEtran}
\IEEEoverridecommandlockouts
\usepackage[utf8]{inputenc} % needed for UTF-8 source files
\DeclareUnicodeCharacter{2248}{\ensuremath{\approx}}
\usepackage{cite}
\usepackage{amsmath,amssymb,amsfonts}
\usepackage{algorithmic}
\usepackage{graphicx}
\usepackage{textcomp}
\usepackage{subcaption}
\usepackage{array}
\usepackage{multirow}
\usepackage{amsmath}
\usepackage{comment}
\usepackage{xcolor}
\usepackage{hyperref}   % makes references clickable
\usepackage{cleveref}   % makes referencing smarter
\usepackage{titlesec}
\titlespacing*{\subsection}{0pt}{0.5em}{0.3em}
\def\BibTeX{{\rm B\kern-.05em{\sc i\kern-.025em b}\kern-.08em
    T\kern-.1667em\lower.7ex\hbox{E}\kern-.125emX}}
\begin{document}

\title{Machine Learning Power Side-Channel Attack on SNOW-V\\
\thanks{ * These authors contributed equally to this work.

This work was supported in part by the Cyber Security Karnataka (CySecK) initiative, Power Grid Centre of Excellence (PGCoE), and in part by the Department of Science \& Technology (DST), India.} 
}

% \author{\IEEEauthorblockN{Deepak*}
% \IEEEauthorblockA{\textit{Dept. of Electronic Systems Engg} \\
% \textit{Indian Institute of Science}\\
% Bengaluru, India  \\
% deepakbalotia@gmail.com}
% \and
% \IEEEauthorblockN{Rahul Balout *}
% \IEEEauthorblockA{\textit{Dept. of Electronic Systems Engg} \\
% \textit{Indian Institute of Science}\\
% Bengaluru, India \\
% balout.rahul2000@gmail.com}
% \and

% \IEEEauthorblockN{Debayan Das}
% \IEEEauthorblockA{\textit{Dept. of Electronic Systems Engg} \\
% \textit{Indian Institute of Science}\\
% Bengaluru, India  \\
% debayandas@iisc.ac.in}

% }

\author{\IEEEauthorblockA{Deepak \IEEEauthorrefmark{1},
Rahul Balout \IEEEauthorrefmark{1},
Anupam Golder\IEEEauthorrefmark{2},
Suparna Kundu\IEEEauthorrefmark{3},
Angshuman Karmakar\IEEEauthorrefmark{3}\IEEEauthorrefmark{5},
Debayan Das\IEEEauthorrefmark{4}}\\
\IEEEauthorblockA{\IEEEauthorrefmark{1}\IEEEauthorrefmark{4}Indian Institute of Science, Bangalore, India} 
\IEEEauthorblockA{\IEEEauthorrefmark{2}Intel Corporation, USA}
\IEEEauthorblockA{\IEEEauthorrefmark{3}KU Leuven, Belgium}
\IEEEauthorblockA{\IEEEauthorrefmark{5}Indian Institute of Technology, Kanpur, India}
}

\maketitle

\begin{abstract}
This paper demonstrates a power analysis-based Side-Channel Analysis (SCA) attack on the SNOW-V encryption algorithm, which is a 5G mobile communication security standard candidate.  Implemented on an STM32 microcontroller, power traces captured with a ChipWhisperer board were analyzed, with Test Vector Leakage Assessment (TVLA) confirming exploitable leakage. Profiling attacks using Linear Discriminant Analysis (LDA) and Fully Connected Neural Networks (FCN) achieved efficient key recovery, with FCN achieving $>5\times$ lower minimum traces to disclosure (MTD) compared to the state-of-the-art Correlational Power Analysis (CPA) assisted with LDA. The results highlight the vulnerability of SNOW-V to machine learning-based SCA and the need for robust countermeasures.
\begin{comment}
    This paper presents an implementation of a Side-Channel Attack (SCA) on the SNOW-V encryption algorithm using power analysis techniques. A ChipWhisperer board was employed to capture power traces while SNOW-V was executed on an STM32 microcontroller. To evaluate potential vulnerabilities, we conducted a Test Vector Leakage Assessment (TVLA), which confirmed exploitable leakage in the current implementation. We then leveraged profiling attacks utilizing both Linear Discriminant Analysis (LDA) and Fully Connected Neural Networks (FCN) to extract the secret key. These machine learning models were trained to classify key-dependent variations in power traces under a known IV setting, enabling efficient and accurate key recovery. The FCN model, in particular, demonstrated strong performance on high-bit classification tasks due to its ability to learn complex non-linear leakage patterns. Our results highlight the vulnerability of the current SNOW-V implementation to side-channel attacks and demonstrate the effectiveness of machine learning-based key recovery techniques, underscoring the need for robust countermeasures to mitigate such threats.
\end{comment}

\end{abstract}

\begin{IEEEkeywords}
SNOW-V, Side-Channel Analysis (SCA), Power Analysis, Profiling Attack, Linear Discriminant Analysis (LDA), Fully Connected Neural Network (FCN), Machine Learning.
\end{IEEEkeywords}

%%%%%%%%%%%%%%%%%%%%%%%%%%%%%%%%%%%%%%%%%%%%%%%%%%%%%%

\section{Introduction}
SNOW-V is a high-performance stream cipher for 5G and beyond, evolved from the SNOW family ~\cite{ekdahl2019snowv} to improve security and efficiency in software environments. It uses a Linear Feedback Shift Register (LFSR) and a Finite State Machine (FSM) to produce a keystream, leveraging SIMD acceleration for high throughput \cite{ekdahl2019snowv}. Although resistant to classical cryptanalysis, SNOW-V has been shown to be susceptible to power Side-Channel Analysis (SCA) attack using a combination of Correlational Power Analysis (CPA) and machine learning (ML) ~\cite{10545384, saurabh2024full}. In general, Machine learning algorithms \cite{chari2002template} such as Linear Discriminant Analysis (LDA) and Fully Connected Neural Networks (FCNs) have shown effectiveness in key recovery against various cryptographic algorithms. In this work, we perform a power SCA attack \cite{kumar2022side} on SNOW-V using a ChipWhisperer and STM32, applying profiling analysis \cite{picek2019profiling} with LDA as well as FCN. %Measurement results show that FCNs achieve superior multi-bit recovery performance, underscoring the need for stronger countermeasures.

\subsection{Motivation}

The first SCA attack on the SNOW-V stream cipher was reported in~\cite{10545384}, combining non-profiling CPA with a profiling refinement using LDA. While CPA identified potential key candidates, it was hindered by ambiguities, such as ghost peaks from nonlinear operations (e.g., \texttt{mul\_x\_inv()}), which required an LDA-based classifier to isolate correct guesses.
Although effective under controlled conditions, this hybrid method still relied on non-profiling steps and required multiple traces for optimal performance.

This paper presents a complete profiling attack using supervised ML models, specifically LDA and FCN, to recover keys directly from the labeled power traces. FCNs capture complex non-linear leakage patterns without manual feature engineering, enabling higher accuracy even under noisy conditions.
 
This work moves toward end-to-end profiling attacks on SNOW-V, reducing reliance on statistical correlation methods like CPA and improving side-channel key extraction efficiency in terms of the minimum traces required to reveal the secret key (Minimum Traces to Disclosure (MTD)).
\subsection{Contribution}
This work presents a Machine Learning-based profiled attack on the SNOW-V cipher, focusing on incremental and high-accuracy key recovery using both classical ML (LDA) and deep learning (FCN) techniques. Our key contributions are as follows.

\begin{itemize}
\vspace{-1mm}
\item We systematically evaluate the effectiveness of SCA on SNOW-V using two profiling-based machine learning techniques: LDA and FCN. LDA is used for its efficiency in low-dimensional classification and ghost peak resolution, while FCNs enable learning complex, non-linear leakage patterns. 

\item We demonstrate progressive recovery of key bits starting with a single bit and extending gradually to 2, 4, and 8 bits by increasing the model complexity.

\item Building on this, we show that deep learning models like FCNs can outperform traditional techniques by achieving high classification accuracy even in the 8-bit secret key recovery. Our approach underscores the strength of profiling attacks using ML and highlights the necessity to implement robust countermeasures, such as masking or hiding, to secure SNOW-V against such attacks.

%\item Our findings provide valuable information on the vulnerability of SNOW-V to advanced profiling attacks, emphasizing the importance of secure and side-channel-resistant implementations in cryptographic systems.

\end{itemize}

%%%%%%%%%%%%%%%%%%%%%%%%%%%%%%%%%%%%%%%%%%%%%%%%%%%%%%%%%%%%%%%%%%%%%%%%%%%%%

\section{BACKGROUND AND RELATED WORK}
\subsection{Evolution of SNOW in Telecommunication} 
The SNOW family of stream ciphers has evolved alongside mobile communication standards. SNOW 1.0 was replaced by SNOW 2.0 for improved security and efficiency, and SNOW 3G became a standardized algorithm for 3G and 4G LTE. To address the performance and security demands of 5G, SNOW-V was introduced~\cite{3GPPSS3-211407, 3GPPTR33.841-2022, ekdahl2019snowv, yang_overview_2020}, optimized for software encryption with SIMD operations, wider registers, and stronger resistance to algebraic attacks. However, its susceptibility to SCA attacks remains largely unexplored, warranting further analysis.

\begin{figure}[!t]
\centerline{\includegraphics[width=0.8\linewidth]{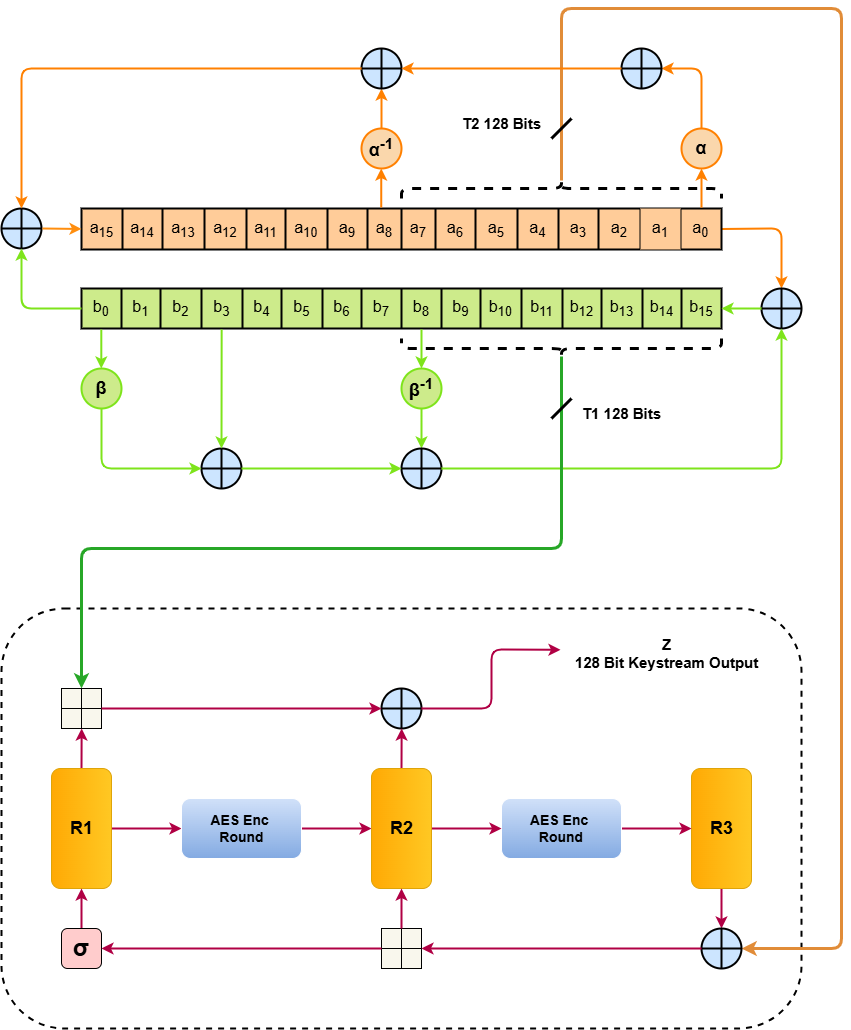}}
\caption{Architecture of SNOW-V, comprising: (a) two LFSRs, each consisting of 16 blocks of 16 bits; (b) multiplication units \texttt{mul\_x}, $\alpha$, $\beta$, and their respective inverses $\alpha^{-1}$ and $\beta^{-1}$; (c) three 128-bit register blocks; and (d) two AES rounds using a round key of $0^{128}$.}
\label{fig:architecture}
\vspace{-2mm}
\end{figure}

\subsection{Architecture of SNOW-V}

\begin{figure}[!t]
    \centering
    
    % First image
    \begin{subfigure}[b]{0.2\textwidth}
        \centering
        \caption{}    
        \includegraphics[width=\linewidth]{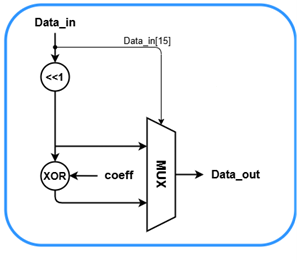}
        
        \label{fig:alpha_beta}
    \end{subfigure}
    \hfill
    % Second image
     \begin{subfigure}[b]{0.2\textwidth}
        \caption{}
        \centering
        \includegraphics[width=\linewidth]{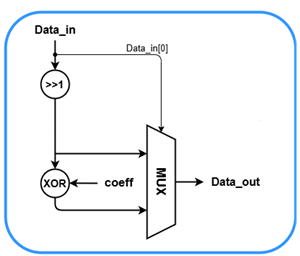}
         
        \label{fig:alpha_beta}
    \end{subfigure}

    \caption{Internal architecture of $\alpha$, $\beta$, $\alpha^{-1}$, and $\beta^{-1}$ : (a) $\alpha$, $\beta$: these are $\texttt{mul\_x}$ structures with coefficients $0 \  X \ 990f$ and $0 \ X \  c963$ for $\alpha$ and $\beta$ respectively; (b) $\alpha^{-1}$, and $\beta^{-1}$: these are $\texttt{mul\_x\_inverse}$ structures with coefficients $0 \ X \  cc87$ and $0\  X \ e4b1$ for $\alpha^{-1}$ and $\beta^{-1}$, respectively.}
    \label{fig:combined_alpha_aes}
    \vspace{-4mm}
\end{figure}

SNOW-V ~\cite{ekdahl2019snowv, caforio_melting_2022} is a high-performance stream cipher designed for 5G security applications, retaining the general structure of its predecessors but introducing enhancements for improved security and efficiency as shown in \cref{fig:architecture} and~\ref{fig:combined_alpha_aes}. It consists of two main components:

\begin{enumerate}
    \item \textbf{LFSR-A and LFSR-B:} Each operates on 256 bits word, with 16 elements in $GF(2^{16})$~\cite{belaid2015ffm}, ensuring a long period and strong diffusion. LFSR-A, with elements $a_{15},...,a_0$, updates as:
    \[
    a_{15}^{(t+1)} = b_0^{(t)} + \alpha a_0^{(t)} + a_1^{(t)} + \alpha^{-1} a_8^{(t)} \mod g_A(\alpha)
    \]
    where $g_A(x) = x^{16} + x^{15} + x^{12} + x^{11} + x^{8} + x^{3} + x^{2} + x + 1 \in GF(2)[x]$.  
    LFSR-B, with elements $b_{15},...,b_0$, updates as:
    \[
    b_{15}^{(t+1)} = a_0^{(t)} + \beta b_0^{(t)} + b_3^{(t)} + \beta^{-1} b_8^{(t)} \mod g_B(\beta)
    \]
    where $g_B(x) = x^{16} + x^{15} + x^{14} + x^{11} + x^{8} + x^{6} + x^{5} + x + 1 \in GF(2)[x]$.  
    Each iteration updates both LFSRs eight times, refreshing 256 bits of the state. Taps $T_1$ and $T_2$ are extracted from updated LFSR segments to feed the FSM.
    
    \item \textbf{FSM:} Consists of three 128-bit registers ($R_1, R_2, R_3$) and generates a 128-bit keystream per update. It uses taps $T_1$ and $T_2$, two fixed-key AES rounds ($C_1 = C_2 = 0^{128}$), modular addition $\boxplus$  on 32-bit subwords, and AES transformations. The LFSRs update eight times before each FSM update to ensure fresh inputs.
\end{enumerate}

SNOW-V employs a 256-bit key and 128-bit IV, and its use of SIMD-based vector operations enables high-speed software implementations. However, side-channel vulnerabilities remain a concern in embedded platforms.

\vspace{-2mm}
\begin{equation}
\sigma = [0, 4, 8, 12, 1, 5, 9, 13, 2, 6, 10, 14, 3, 7, 11, 15]
\label{eq:sigma}
\end{equation}

The $\sigma$ permutation maps byte $i$ to its new position (byte 0 → pos 0, byte 4 → pos 1, etc.), implementing a column-wise transposition of the AES state matrix.

\subsection{Profiling-based ML SCA Using LDA}
Profiling SCA utilizes machine learning techniques to exploit power consumption patterns for key recovery. LDA is effective for this task, reducing dimensionality while preserving key-dependent class separability. Compared to CPA, LDA is more robust to noise and often requires fewer traces \cite{choudary_efficient_2014}. While used in prior SCA research, its application to SNOW-V is largely unexplored. Here, we employ a purely LDA-based profiling approach for bitwise key recovery, starting from the least significant bits and progressively reconstructing the full key.

\subsection{Deep Learning-Based SCA Using FCN}
To improve profiling performance, we also employ FCN, which can model complex non-linear relationships in power trace data \cite{das2019xdeepsca}. Traces are first processed with Principal Component Analysis (PCA) to reduce dimensionality and noise. Using an identity model for labeling \cite{das2019xdeepsca}, the FCN consistently outperforms LDA, especially in recovering multiple bits together, showing a strong generalization for single-trace recovery. This demonstrates the practicality of deep learning-based SCA on SNOW-V.

\section{ATTACK METHODOLOGY}

\subsection{Initial Findings}
Our analysis of SNOW-V reveals potential SCA attack targets:
\begin{itemize}
\item \textbf{LFSR:} Stores the secret key and IV during initialization, making it the most susceptible component.
\item \textbf{FSM Transitions:} State changes may leak key-related information.
\item \textbf{Field Multiplication $(mul_x)$:} Certain key scheduling operations may contribute to leakage.The internal structure is shown in \cref{fig:combined_alpha_aes}.
\end{itemize}
During initialization, the LFSR holds the complete key and IV, while the FSM registers and AES round keys (C1, C2) are set to zero. Consequently, the AES unit does not exhibit key-dependent leakage, serving only to scramble the output stream.
\begin{comment}
Our analysis of the SNOW-V cipher highlights several potential vulnerabilities that can be exploited in a Side-Channel Attack (SCA):

\begin{itemize}
    \item Linear Feedback Shift Register (LFSR): The structured nature of its operation may inadvertently expose key-dependent patterns.
    \item Finite State Machine (FSM) Transitions: Changes in the internal state could be leveraged to infer sensitive information.
    \item Field Multiplication $(mul\_x)$: Weaknesses in key scheduling operations may contribute to key leakage.

\end{itemize}

Among these, the LFSR is particularly susceptible, as it is responsible for storing the secret key and initialization vector (IV) during the setup phase.

During initialization, the LFSR is populated with the complete key and IV, while the registers in the FSM are set to zero. Additionally, the round keys C1 and C2 used in the AES rounds within the FSM are also initialized to zero. As a result, the AES component becomes an unsuitable target for an attack, as it does not contribute to key dependency and is solely responsible for scrambling the output stream.

\end{comment}

According to the SNOW-V specification \cite{ekdahl2019snowv}, the initialization process follows the mappings shown in \cref{tab:lfsr_init}.

\begin{table}[!t]
\centering
\renewcommand{\arraystretch}{1.5}

% First LFSR Table
\begin{tabular}{|c|c|}
\hline
\textbf{LFSR A} & $A_1, A_1, A_2, \ldots, A_6, A_7,\quad A_8, A_9, \ldots, A_{14}, A_{15}$ \\
\hline
\textbf{Initial Value} & $IV_0, IV_1, \ldots, IV_6, IV_7,\quad K_{00}, K_{01}, \ldots, K_{06}, K_{07}$ \\
\hline
\end{tabular}
% qqqqqq
\vspace{1em}

% Second LFSR Table
\begin{tabular}{|c|c|}
\hline
\textbf{LFSR B} & $B_0, B_1, B_2, \ldots, B_6, B_7,\quad B_8, B_9, \ldots, B_{14}, B_{15}$ \\
\hline
\textbf{Initial Value} & $0x0, 0x0, \ldots, 0x0, 0x0,\quad K_{08}, K_{09}, \ldots, K_{14}, K_{15}$ \\
\hline
\end{tabular}

\caption{Initialisation of LFSR: LFSR A is half filled with the entire 128-bit IV and the other half with the first half of the 256-bit key. LFSR B is half filled with the remaining part of the key and the other half with zeros.}
\label{tab:lfsr_init}
\vspace{-6mm}
\end{table}

Here, the secret key K is represented as $(k_{15}, k_{14},...,k_{1},k_{0}$ while the IV is given by $(iv_{7}, iv_{6},...,iv_{1},iv_{0})$.
Each key byte $k_i$ and $IV$ byte $iv_j$ (for $0 \leq i \leq 15$ and $0 \leq j \leq 7$ ) consists of a 16-bit value.

These equations indicate that the LFSR, particularly the sections holding the key, is a critical attack target. 
%%%%%%%%%%%%%%%%%%%%%%%%%%%%%%%%%%%%%%%%%%%%%%%%%%%%%%%%%%%%%%%%%%%%%%%%%%%%%%%%%%%%%%%%%%

\subsection{Strategy for Side Channel Attack on SNOW-V}
The LFSRs serve as the primary target for key extraction due to their role in storing the 256-bit secret key. In the initialization state, the most significant 128 bits are stored in LFSR B (b[15]--b[8]) and the least significant 128 bits in LFSR A (a[15]--a[8]). The known 128-bit IV is loaded into LFSR A (a[7]--a[0]), while the lower half of LFSR B is set to zero, as shown in \cref{tab:lfsr_init}.

\begin{figure}[!t]

    \centering
    \includegraphics[width=1\linewidth]{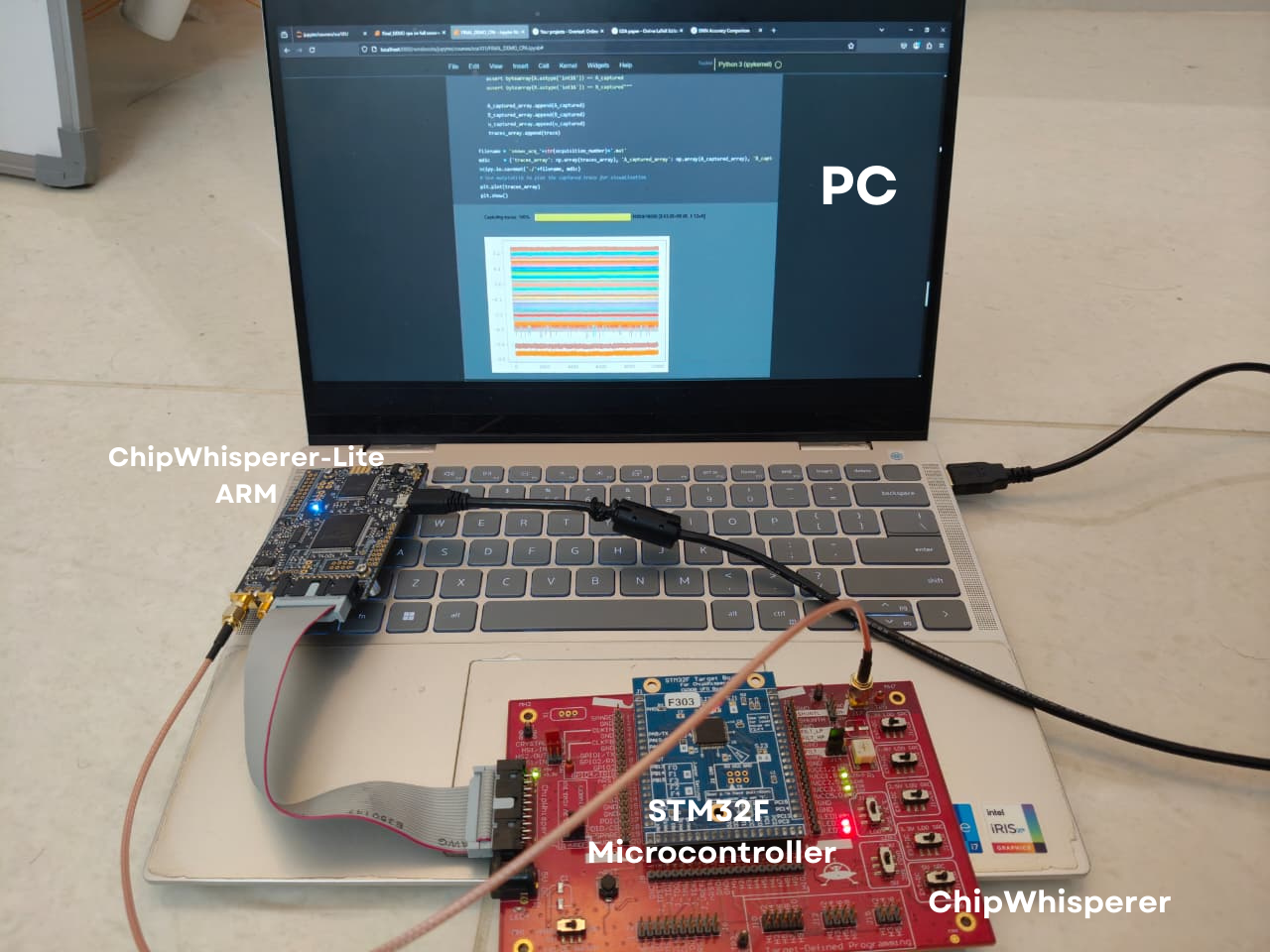}
    \caption{Hardware setup showing workstation, ChipWhisperer capture and target boards, and an STM32 embedded board mounted on the target board and connected via cables. The STM32 microcontroller operates at a clock frequency of 7.37 MHz and sampling rate is four times the clock frequency.}
    \label{fig:hardware_setup}
\end{figure}

\begin{figure*}[t]
    \centering
    \begin{subfigure}[t]{0.48\textwidth}
        \centering
        \caption{} 
    \includegraphics[width=1\linewidth]{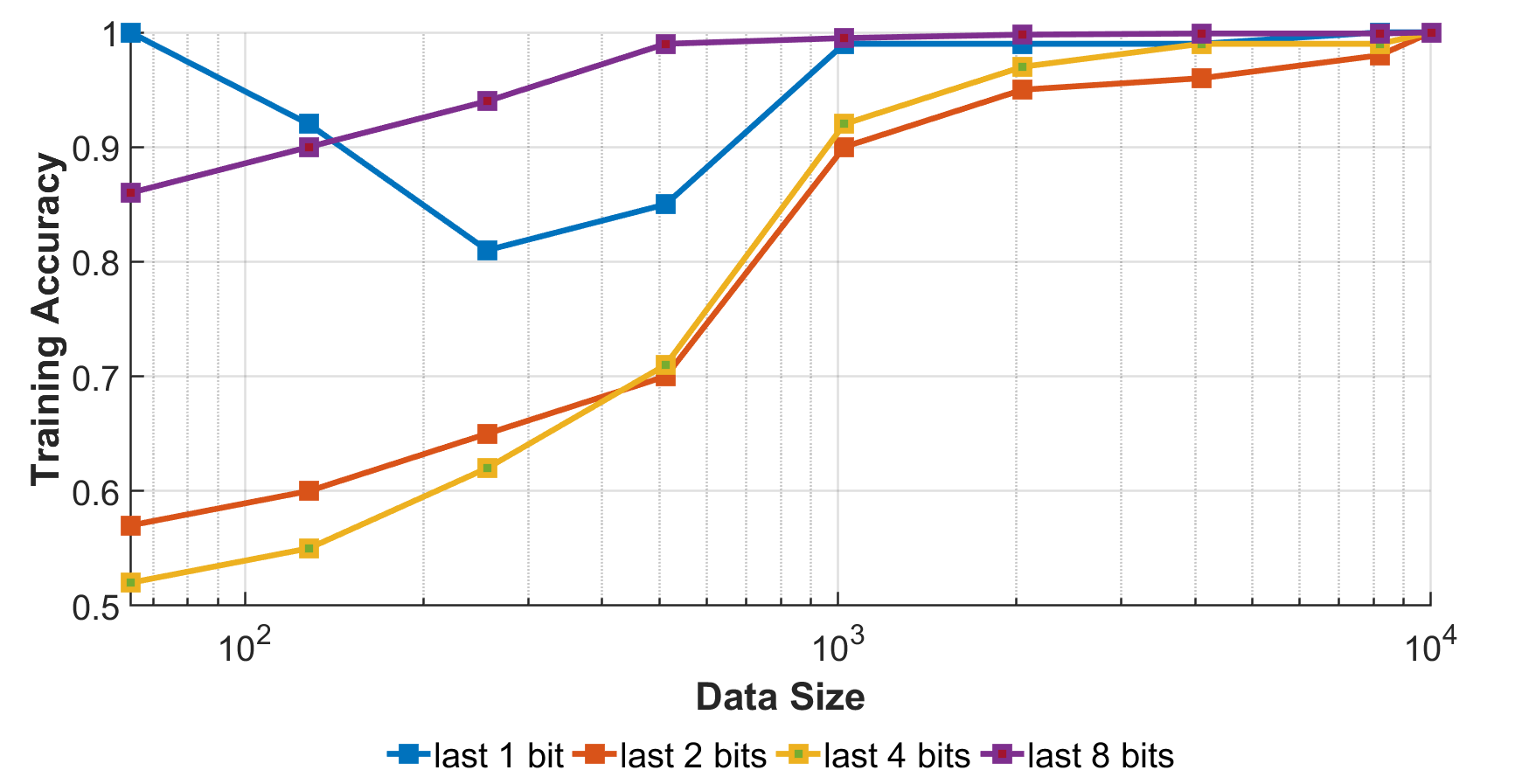}

        \label{fig:lda-testing}
    \end{subfigure}
    \hfill
    \begin{subfigure}[t]{0.48\textwidth}
    \caption{} 
        \centering
    \includegraphics[width=1\linewidth]{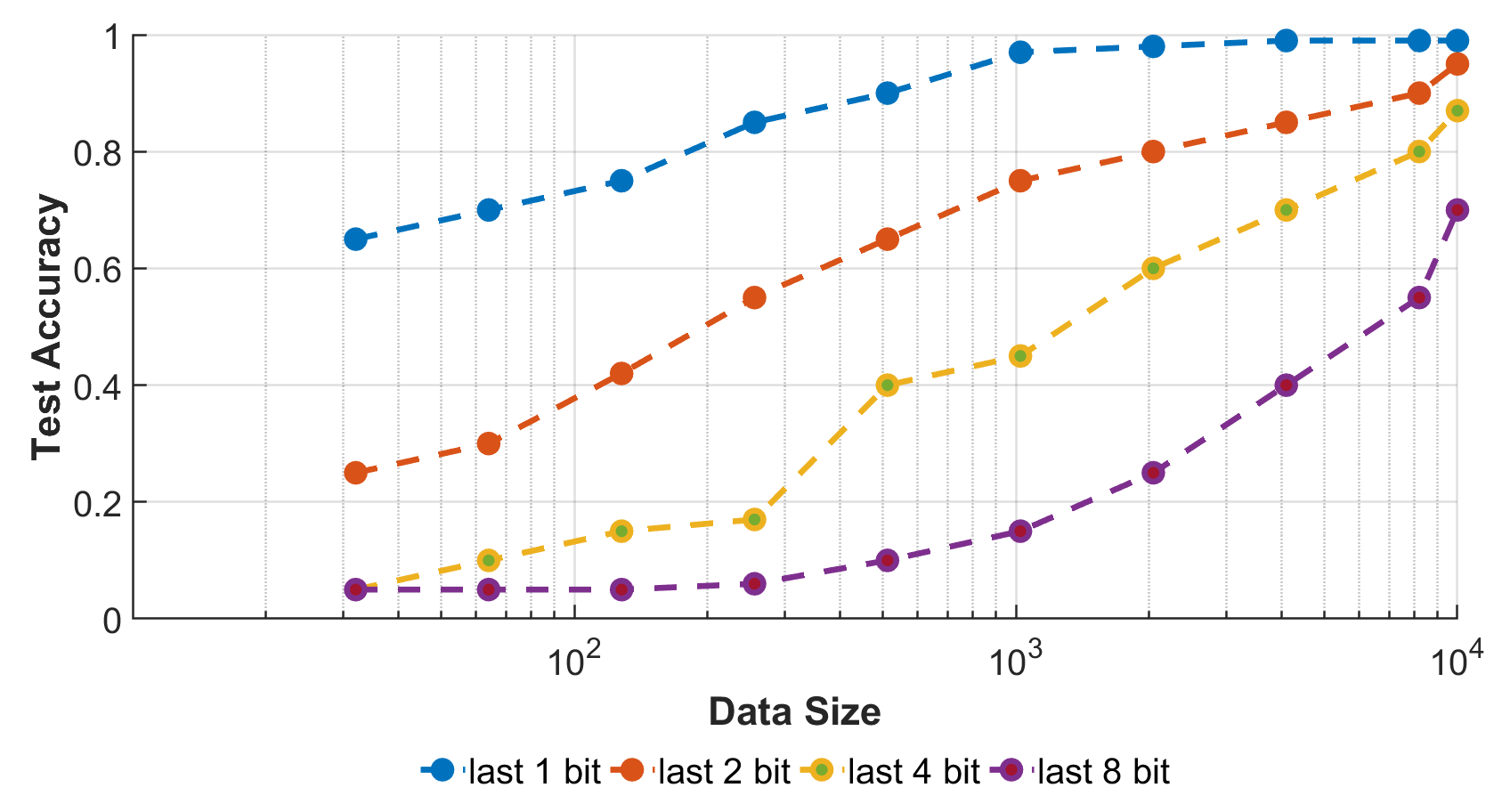}

        \label{fig:lda-training}
    \end{subfigure}
    \vspace{-2mm}
    \caption{ LDA classification accuracy for recovering bits of the internal state word $A[8]$ across varying data sizes with a 64:16:20 training–validation–testing split. (a) Training accuracy for the last $n$ bits ($n=1,2,4,8$), showing improvement as training traces increase. (b) Test accuracy for the last $n$ bits, demonstrating reliable recovery even for 8-bit cases with increasing traces.}

    \label{fig:lda-combined}
    \vspace{-4mm}
\end{figure*}

Before entering the FSM stage, the LFSRs perform eight update iterations. We focus on the update functions for registers a[15] and b[15], expressed as:
\[
u = mul_x (A[0], 0x990f) \oplus A[1] \oplus mul_x^{inv} (A[8], 0xcc87) \oplus B[0]
\]
\[
v = mul_x (B[0], 0xc963) \oplus B[3] \oplus mul_x^{inv} (B[8], 0xe4b1) \oplus A[0]
\]

In the first iteration, $u$ depends on $A[0]$, $A[1]$, $A[8]$, and $B[0]$. Since $A[0]$, $A[1]$, and $B[0]$ are known, $A[8]$ can be directly recovered ~\cite{saurabh2024full}. This process is repeated for $u$ and $v$ in subsequent iterations, each time revealing new key-related LFSR blocks. By systematically applying this method, all segments of the secret key can be extracted.

\subsection{Attack Algorithms}

The SCA begins with a \textbf{TVLA}~\cite{schneider2015leakage} using 100{,}000 power traces captured under fixed and random IV conditions to identify the potential source of the leakage within the algorithm. Leakage points are identified where the absolute $t$-value exceeds \textbf{4.5}~\cite{schneider2015leakage}, confirming exploitable data-dependent leakage. %Trigger points are inserted at LFSR update locations to capture key-dependent operations.

Two profiling attack techniques are applied:

\begin{itemize}
    \item \textbf{LDA-Based Attack:} Supervised LDA models are trained on traces labeled with the last 1, 2, 4, or 8 bits of the target key byte. LDA reduces dimensionality while maximizing class separability. The training and testing set is split as 80:20, and performance is evaluated with 10{,}000, 50{,}000, and 100{,}000 training traces as discussed in Section IV.
    
    \item \textbf{FCN-Based Attack:} FCN model the non-linear relationship between leakage and key values. Power traces are preprocessed using PCA to the top 2{,}000 components (preserving 99\% variance) and Known Value Correlation (KVC) to isolate highly key-dependent segments, typically during early LFSR updates. The FCN architecture has three dense hidden layers (512, 256, 128 neurons) with ReLU, LeakyReLU, PReLU, ELU, or SELU activations, followed by a softmax output layer. Models are trained for 100 epochs on the PCA-reduced, KVC-filtered dataset, achieving high accuracy in multi-bit key recovery as discussed in the upcoming section.
\end{itemize}

\begin{figure*}[t]
    \centering

    % First row
    \begin{subfigure}[t]{0.48\textwidth}
        \centering
        \caption{}
        \includegraphics[width=\linewidth]{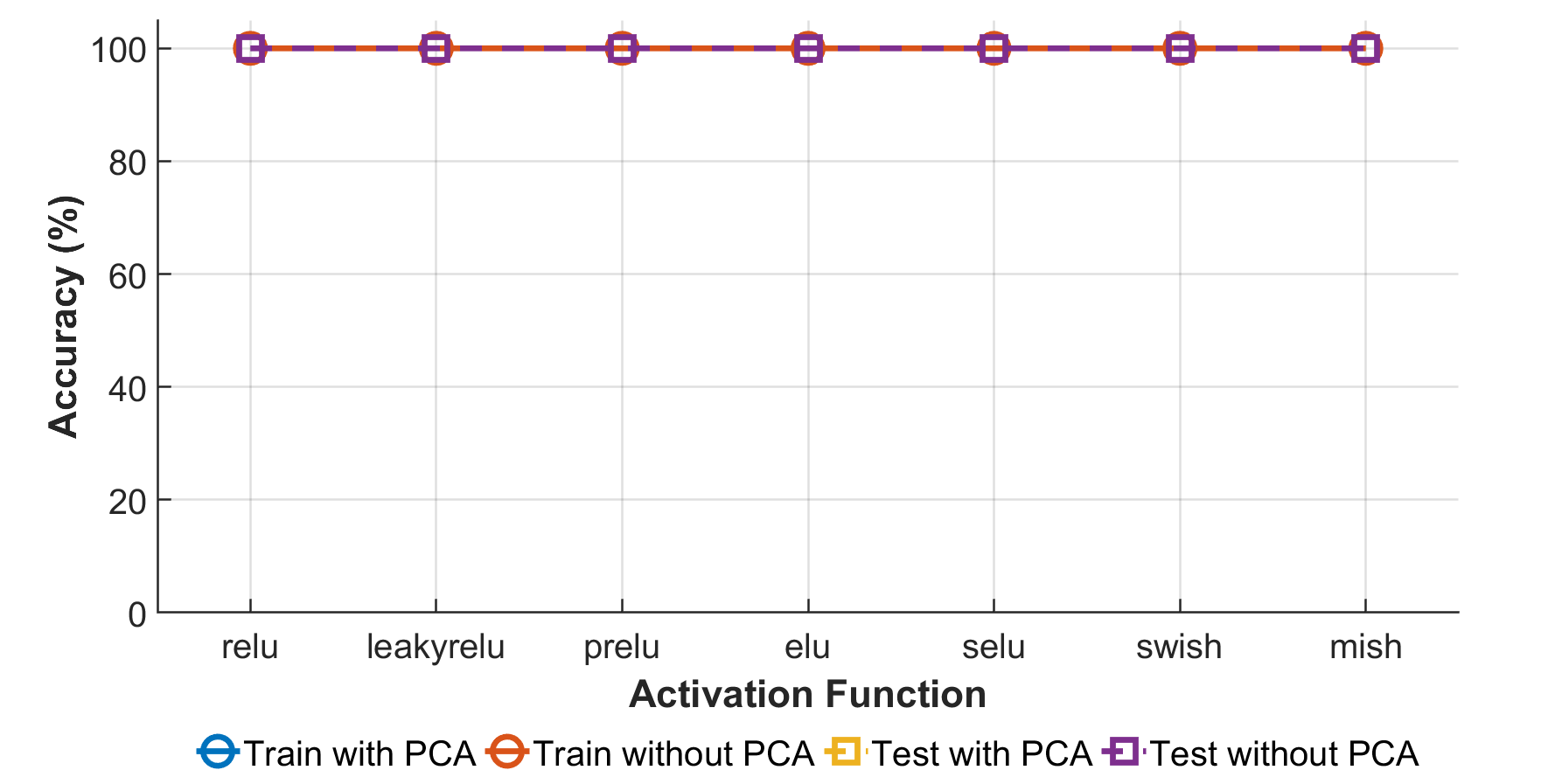}

        \label{fig:fcn-1bit}
    \end{subfigure}
    \hfill
    \begin{subfigure}[t]{0.48\textwidth}
        \centering
        \caption{}
        \includegraphics[width=\linewidth]{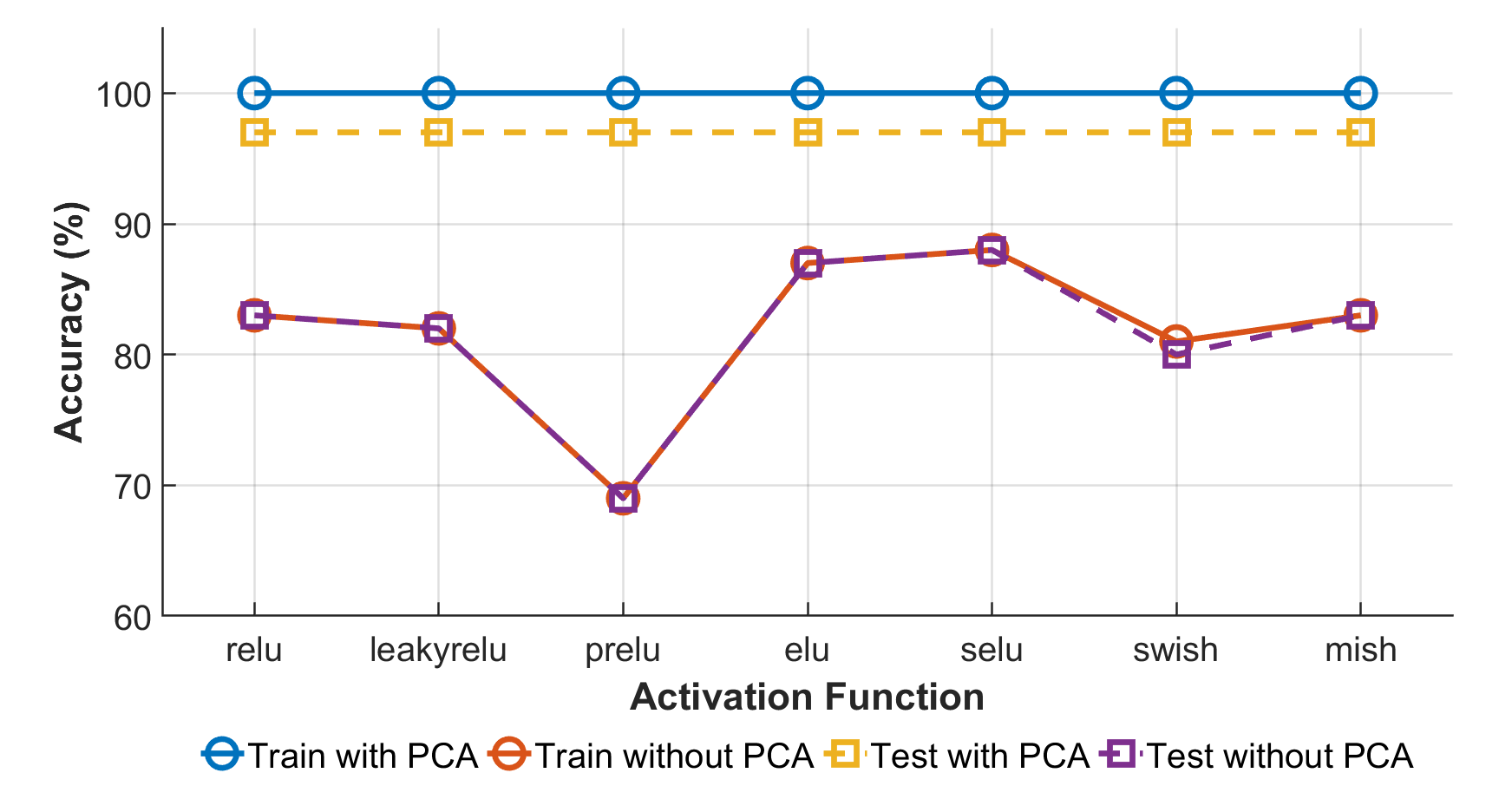}

        \label{fig:fcn-2bit}
    \end{subfigure}

    \vspace{0.5em} % Optional vertical spacing between rows

    % Second row
    \begin{subfigure}[t]{0.48\textwidth}
        \centering
        \caption{}
        \includegraphics[width=\linewidth]{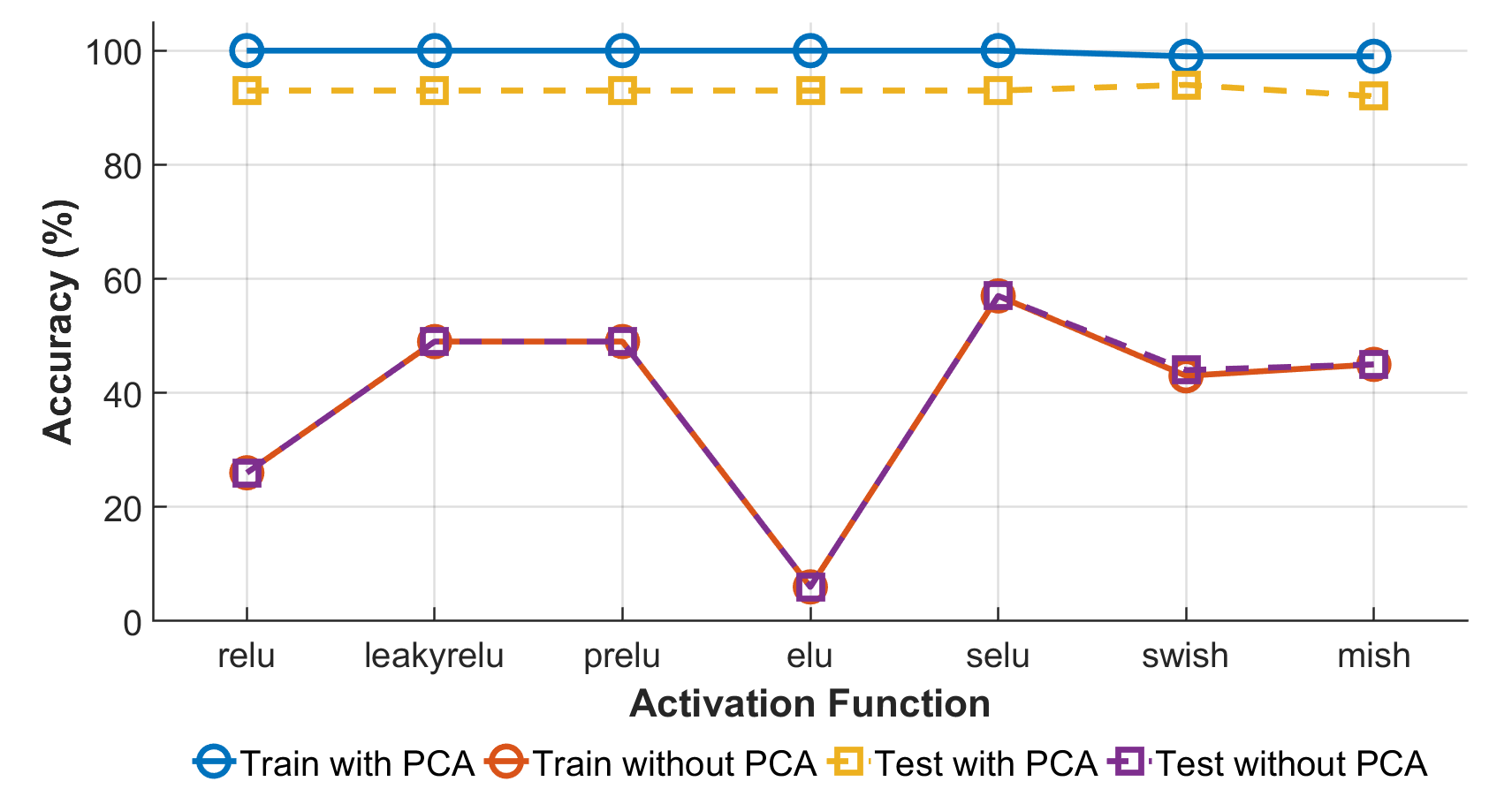}

        \label{fig:fcn-4bit}
    \end{subfigure}
    \hfill
    \begin{subfigure}[t]{0.48\textwidth}
        \centering
        \caption{}
        \includegraphics[width=\linewidth]{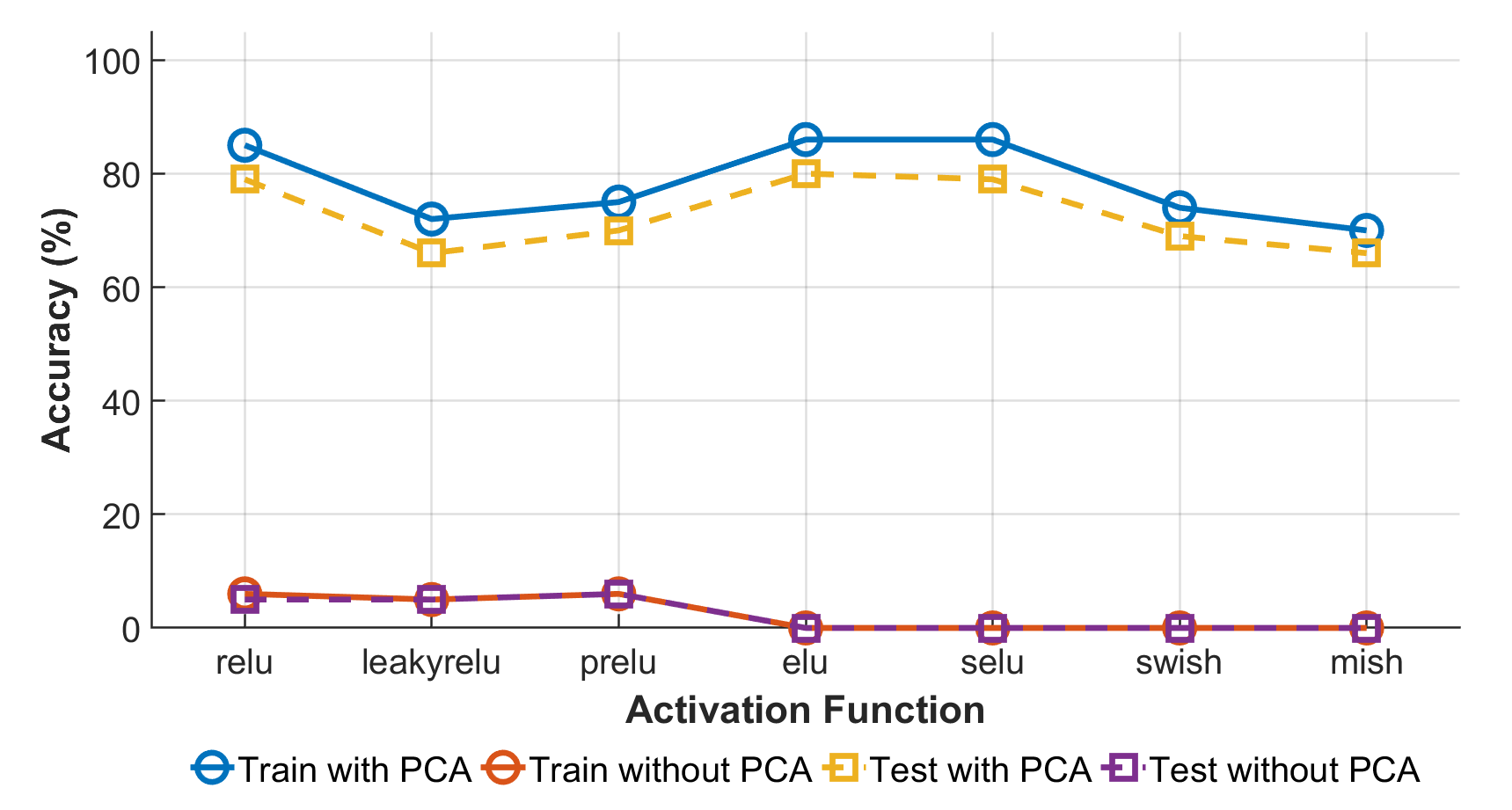}

        \label{fig:fcn-8bit}
    \end{subfigure}
    %\vspace{-2mm}
    \caption{   Accuracy of bit recovery for the internal state word $A[8]$ using a FCN trained on $10^5$ traces, with and without PCA (80:20 train–test split). (a)~1-bit recovery shows negligible PCA impact; (b)~2-bit recovery improves with PCA, especially for PReLU layers; (c)~4-bit recovery drops overall, but PCA gives up to ~15\% gain; (d)~8-bit recovery falls sharply, with non-PCA models under 10\%.  }
    \vspace{-2mm}
    %Accuracy of bit recovery for the internal state word $A[8]$ using a Fully Connected Network (FCN) model trained on $10^5$ side-channel traces, with and without the application of Principal Component Analysis (PCA). Out of the total traces, 80\% were used for training the model, and the remaining 20\% were reserved for testing. The comparison is conducted across different network layers to evaluate the impact of dimensionality reduction on learning performance. : (a) Training and testing accuracy for 1-bit recovery of $A[8]$ using an FCN model on $10^5$ traces, with and without PCA. PCA has negligible impact, indicating dimensionality reduction does not affect single-bit classification accuracy; (b) 2-bit recovery accuracy for $A[8]$ comparing FCN models with and without PCA. PCA yields better overall results; without PCA, PReLU-based layers underperform compared to other activations; (c) 4-bit recovery accuracy for $A[8]$. PCA models outperform non-PCA ones, but accuracy drops compared to the 2-bit case. Non-PCA performance degrades sharply, underscoring PCA’s role as recovery complexity increases; (d)  8-bit recovery accuracy for $A[8]$. Both PCA and non-PCA accuracies fall markedly; non-PCA achieves under 10\%, showing the challenge of high-bit recovery and the limits of both PCA and raw feature learning.}

    \label{fig:fcn-multi}
\end{figure*}
 %%%%%%%%%%%%%%%%%%%%%%%%%%%%%%%%%%%%%%%%%%%%%%%%%%%%%%%%%%%%%%%%

\section{Experimental Results}
%\vspace{-1mm}
To assess the effectiveness of profiling side-channel attacks on the SNOW-V stream cipher, we evaluated two machine learning-based classifiers: \textbf{LDA} and \textbf{FCN}. The experiments were conducted using power traces captured from an STM32 microcontroller as shown in \cref{fig:hardware_setup}, with the goal of recovering the last 1, 2, 4, and 8 bits of a target key byte.

\subsection{LDA-Based Classification}

LDA showed strong performance in low-dimensional classification tasks. The training and testing accuracy are shown in \cref{fig:lda-combined}. For the 1-bit recovery scenario, it achieved an accuracy of approximately \textbf{99\%}, and for 2-bit recovery, accuracy ranged between \textbf{90\% and 95\%}. However, as the number of target key bits increased, classification accuracy declined. For the 8-bit recovery task, the accuracy dropped to around \textbf{60\%}, indicating LDA's limited capacity to handle complex or non-linear leakage. Fig.~\ref{fig:lda-combined} further shows that LDA achieves over \textbf{90\%} accuracy for 1- and 2-bit recovery with as few as 1{,}000 traces, while 4- and 8-bit recovery require larger datasets. With 100{,}000 traces, the model exceeds \textbf{95\%} accuracy even for 8-bit classification of $A[8]$. To improve the attack success in such cases, \textbf{majority voting} across multiple traces was applied~\cite{das2019xdeepsca}, which provided some improvement but did not fully compensate for the underlying limitations.

\subsection{FCN-Based Classification}
The FCN-based approach significantly outperformed LDA across all bit-recovery configurations. The accuracies across different layers and the corresponding bit recovery configurations are presented in \cref{fig:fcn-multi}. For 1-bit classification as shown in \cref{fig:fcn-1bit}, the FCN achieved near-perfect accuracy (\textbf{$\sim100\%$}), and it maintained high accuracy even in the more challenging 8-bit scenario ($\sim \textbf{80\%}$) in \cref{fig:fcn-8bit}. This demonstrates the FCN’s ability to learn non-linear relationships and extract meaningful features from high-dimensional trace data without requiring manual feature selection. Despite the increased computational cost and training time, the FCN provided more robust and consistent results across all test cases.
Furthermore, the effect of PCA was analyzed by training FCN models with and without dimensionality reduction. For 1-bit recovery in \cref{fig:fcn-1bit}, PCA had a negligible impact, with both models achieving $\sim \textbf{100\%}$ accuracy. However, as recovery complexity increased, PCA-based models consistently outperformed non-PCA ones. For 2-bit recovery in \cref{fig:fcn-2bit}, PCA improved accuracy from around \textbf{85\%} (non-PCA) to \textbf{95\%}, while for 4-bit recovery \cref{fig:fcn-4bit} it boosted accuracy from nearly \textbf{50\%} to above \textbf{75\%}. Most notably, in the 8-bit case \cref{fig:fcn-8bit}, PCA-based models retained about \textbf{80\%} accuracy, whereas non-PCA models dropped below \textbf{10\%}. These results highlight the critical role of dimensionality reduction in stabilizing FCN performance for higher-bit recovery.

%The FCN-based approach significantly outperformed LDA across all bit-recovery configurations. The accuracies across different layers and the corresponding bit recovery configurations are presented in \cref{fig:fcn-multi}.  For 1-bit classification as shown in \cref{fig:fcn-1bit}, the FCN achieved near-perfect accuracy (\textbf{$\sim100\%$}), and it maintained high accuracy even in the more challenging 8-bit scenario (approximately \textbf{80\%}) in \cref{fig:fcn-8bit}. This demonstrates the FCN’s ability to learn non-linear relationships and extract meaningful features from high-dimensional trace data without requiring manual feature selection. Despite the increased computational cost and training time, the FCN provided more robust and consistent results across all test cases.

\subsection{Impact of PCA on FCN Performance}

We further evaluated the performance of FCN without and with \textbf{ (PCA)} applied to the input traces, as shown in  \cref{tab:accuracy_without_pca} and \cref{tab:accuracy_pca}. PCA was used to retain \textbf{99\% of the total variance}, significantly reducing the dimensionality of the input features. This dimensionality reduction led to notable improvements in classification accuracy, particularly in higher-bit recovery tasks. For the 8-bit classification, the use of PCA improved the accuracy, reduced training time, and helped mitigate overfitting by filtering out noisy or irrelevant components from the trace data. The improvement is most evident in the accuracy trends shown in the performance plots, where PCA-enhanced FCN models consistently outperform their non-PCA counterparts.

\begin{table}[!t]
\centering
\caption{Accuracy (\%) with FCN without using PCA for Different Bit Levels and Activation Functions}
\label{tab:accuracy_without_pca}
\renewcommand{\arraystretch}{1.2}
\setlength{\tabcolsep}{4pt} % Slightly increased from 4pt
\footnotesize % Slightly larger than scriptsize for better readability
\begin{tabular}{|c|cc|cc|cc|cc|}
\hline
\textbf{Activation} & \multicolumn{2}{c|}{\textbf{1-bit}} & \multicolumn{2}{c|}{\textbf{2-bit}} & \multicolumn{2}{c|}{\textbf{4-bit}} & \multicolumn{2}{c|}{\textbf{8-bit}} \\
\cline{2-9}
\textbf{Function} & \textbf{Train} & \textbf{Test} & \textbf{Train} & \textbf{Test} & \textbf{Train} & \textbf{Test} & \textbf{Train} & \textbf{Test} \\
\hline
ReLU        & 100 & 100 & 83 & 83 & 26 & 26 & 6 & 5\\
LeakyReLU   & 100 & 100 & 82 & 82 & 49 & 49 & 5 & 5 \\
PReLU       & 100 & 100 & 69 & 69 & 49 & 49 & 6 & 5 \\
ELU         & 100 & 100 & 87 & 87 & 6 & 6 & 0 & 0 \\
SELU        & 100 & 100 & 88 & 88 & 57 & 57 & 0 & 0 \\
Swish       & 100 & 100 & 81 & 80 & 44  & 43 & 0 & 0 \\
Mish        & 100 & 100 & 83 & 83 & 45  & 45 & 0 & 0 \\
\hline
\end{tabular}
\vspace{-2mm}
\end{table}

\begin{table}[ht]
\centering
\caption{Accuracy (\%) using FCN + PCA for Different Bit Levels and Activation Functions}
\label{tab:accuracy_pca}
\renewcommand{\arraystretch}{1.2}
\setlength{\tabcolsep}{4pt} % Slightly increased from 4pt
\footnotesize % Slightly larger than scriptsize for better readability
\begin{tabular}{|c|cc|cc|cc|cc|}
\hline
\textbf{Activation} & \multicolumn{2}{c|}{\textbf{1-bit}} & \multicolumn{2}{c|}{\textbf{2-bit}} & \multicolumn{2}{c|}{\textbf{4-bit}} & \multicolumn{2}{c|}{\textbf{8-bit}} \\
\cline{2-9}
\textbf{Function} & \textbf{Train} & \textbf{Test} & \textbf{Train} & \textbf{Test} & \textbf{Train} & \textbf{Test} & \textbf{Train} & \textbf{Test} \\
\hline
ReLU        & 100 & 100 & 100 & 97 & 100 & 93 & 85 & 79 \\
LeakyReLU   & 100 & 100 & 100 & 97 & 100 & 93 & 72 & 66 \\
PReLU       & 100 & 100 & 100 & 97 & 100 & 93 & 75 & 70 \\
ELU         & 100 & 100 & 100 & 97 & 100 & 93 & 86 & 80 \\
SELU        & 100 & 100 & 100 & 97 & 100 & 93 & 86 & 79 \\
Swish       & 100 & 100 & 100 & 97 & 99  & 94 & 74 & 69 \\
Mish        & 100 & 100 & 100 & 97 & 99  & 92 & 70 & 66 \\
\hline
\end{tabular}
\vspace{-1mm}
\end{table}

\begin{table}[h]
    \centering
    \caption{LDA Classification Accuracy for Different Power Traces}
    \renewcommand{\arraystretch}{1.3}
    \setlength{\tabcolsep}{5pt}

    \begin{tabular}{|c|c|c|c|}
        \hline
        & \textbf{10,000 Traces} & \textbf{50,000 Traces} & \textbf{100,000 Traces} \\
        \hline
        \textbf{Last bit of Key}  & 0.9995  & 1.0     & 1.0     \\
        \textbf{Last 2 bits of Key}  & 0.815   & 0.8965  & 0.95095 \\
        \textbf{Last 4 bits of Key}  & 0.451   & 0.6841  & 0.84815 \\
        \textbf{Last 8 bits of Key}  & 0.102   & 0.3431  & 0.57455 \\
        \hline
    \end{tabular}
    \vspace{-2mm}
     
    \label{tab:lda_accuracy}
\end{table}

%%%%%%%%%%%%%%%%%%%%%%%%%%%%%%%%%%%%%%%%%%%%%%%%%%%%%%%%%%%%%%%%%%%%%

\begin{table}[h]
\centering
\caption{Comparison of Profiling Attacks (LDA vs PCA + FCN) with n bit Accuracy, where n = 1, 2, 4, and 8}
\label{tab:profiling_attack_results}
\renewcommand{\arraystretch}{1.4} % Row height
\setlength{\tabcolsep}{8pt}       % Column padding
\fontsize{7.5pt}{7.5pt}\selectfont % Font size
\begin{tabular}{|l|c|c|}
\hline
\textbf{Aspect} & \textbf{LDA} & \textbf{PCA + FCN} \\
\hline
\textbf{Attack Type} & Profiling & Profiling (deep learning) \\\hline
 \multicolumn{3}{|c|}{Performance}\\\hline
\textbf{1-bit key} & High Accuracy (≈ 99\%) & Perfect Accuracy (≈ 100\%) \\\hline
\textbf{2-bit key} & Good Accuracy (≈ 90–95\%) & High Accuracy (≈ 97–99\%) \\\hline
\textbf{4-bit key} & Good Accuracy (≈ 70–80\%) & Good Accuracy (≈ 90\%) \\\hline
\textbf{8-bit key} & Fair Accuracy (≈ 60\%) & Good Accuracy (≈ 80\%) \\\hline
\textbf{Complexity} & Low & High \\
\hline
\end{tabular}
\vspace{-2mm}
\end{table}

% \subsection{Comparison of LDA vs FCN+PCA attack models on SNOW-V}
% We explored two profiling attack strategies: LDA and FCN. LDA demonstrated effectiveness in low-bit width recovery scenarios, enabling step-by-step key reconstruction. FCN, on the other hand, proved significantly more powerful by capturing complex, non-linear leakage characteristics, thereby achieving higher accuracy across multiple bit widths. Among all activation functions tested, the Exponential Linear Unit (ELU) consistently achieved the highest accuracy for key recovery—including for the last 1, 2, 4, and 8 bits—highlighting FCN with ELU activation as a particularly effective architecture. To achieve high SCA attack success, we performed a majority vote for a given number of traces, as shown in \cref{fig:majority voting}. 

\subsection{Comparison of LDA vs FCN+PCA attack models on SNOW-V}
\begin{figure}[!t]
    \centering
    \includegraphics[width=1\linewidth]{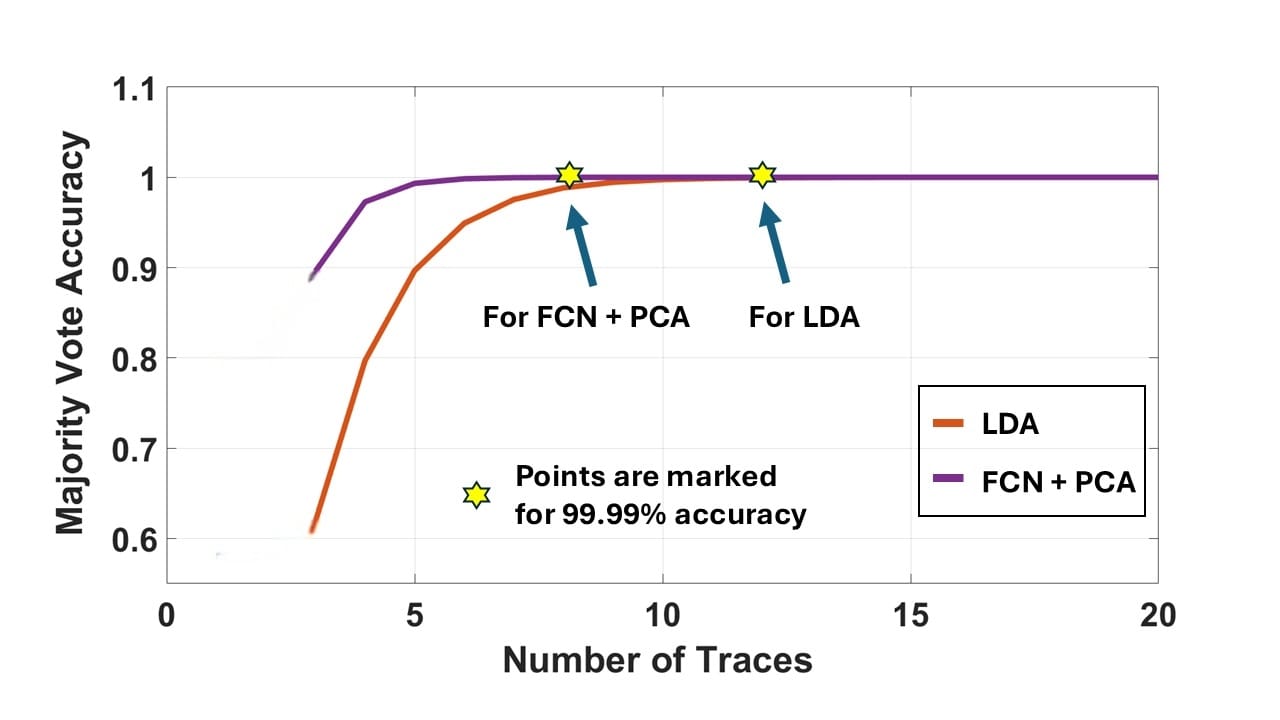}
    \vspace{-8mm}
    \caption{Majority-vote accuracy as a function of the number of traces for: LDA-based analysis and FCN with PCA. Accuracy probability is computed using the majority-voting formula described in \cite{das2019xdeepsca}. The plot shows LDA requires 12 traces, whereas FCN assisted with PCA requires only 8 traces for 99.99\% for majority voting-based accuracy.}
    \label{fig:majority voting}
    \vspace{-4mm}
\end{figure}

    We explored two profiling attack strategies: LDA and FCN. LDA demonstrated effectiveness in low-bit width recovery scenarios, enabling step-by-step key reconstruction. FCN, on the other hand, proved significantly more powerful by capturing complex, non-linear leakage characteristics, thereby achieving higher accuracy across multiple bit widths. We further assessed different activation functions, namely ReLU, Leaky ReLU, PReLU, SELU, and ELU. SELU and ELU consistently achieved the highest accuracy, likely due to their negative and zero-centered activations, which allow the network to better capture subtle trace variations indicative of key leakage~\cite{Dubey2022ActivationFunctions}. ReLU delivered moderate performance, benefiting from suppression of noisy negative values but also discarding potentially informative leakage below zero. Leaky ReLU yielded the lowest accuracy, as its linear negative slope tends to preserve noise, reducing the model’s ability to isolate key-dependent features. To achieve high SCA attack success, we performed a majority vote for a given number of traces, as shown in \cref{fig:majority voting}.

\begin{figure}[!t]
    \centering
    \includegraphics[width=1\linewidth]{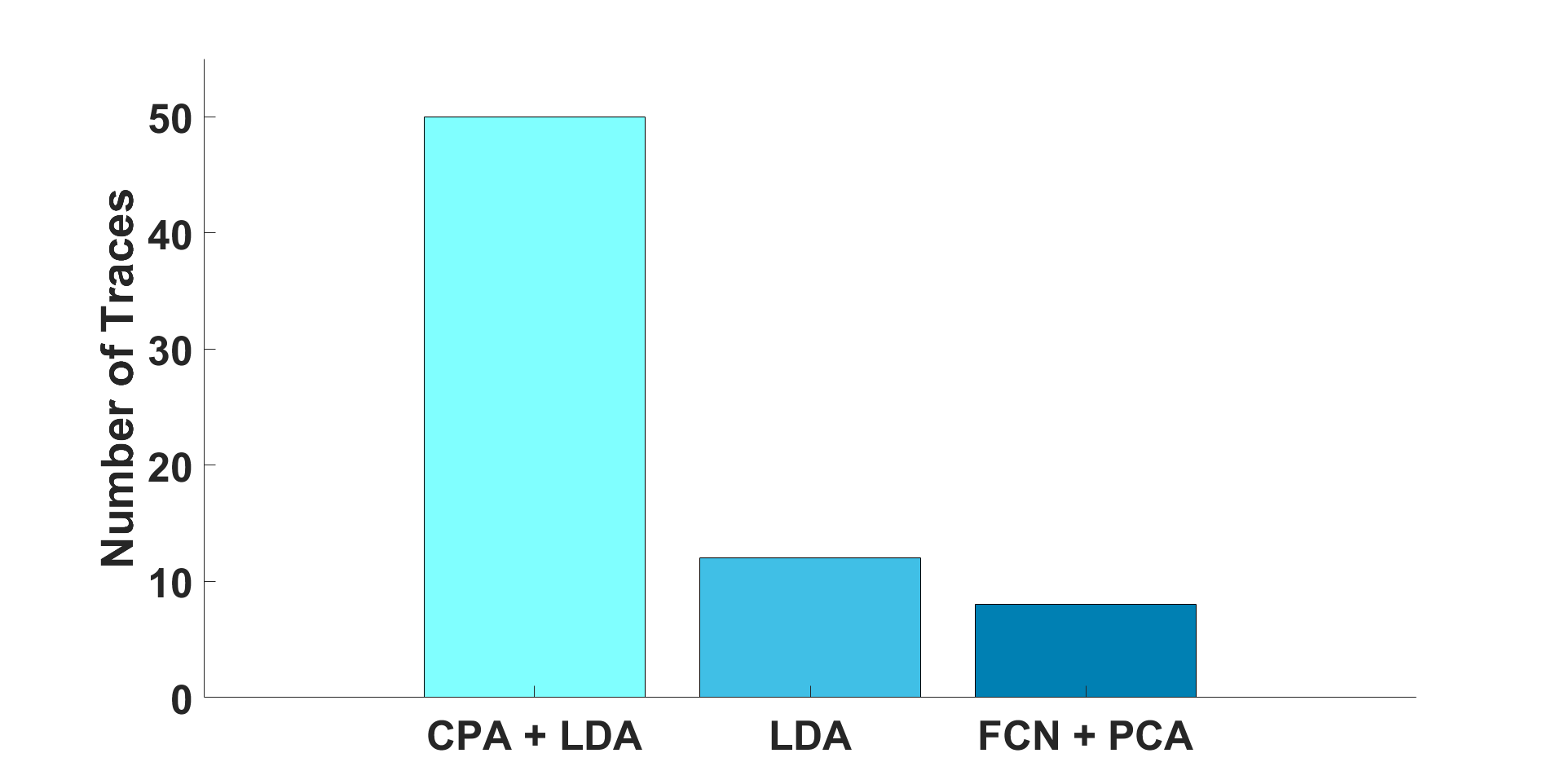}
    \vspace{-2mm}
    \caption{Minimum number of traces required to achieve 99.99\% SCA attack accuracy in recovering the correct key for SNOW-V using three attack methods - CPA+LDA (prior work \cite{10545384}), proposed LDA, and FCN+PCA. }
    % \caption{ Minimum number of traces required to achieve 100\% accuracy in recovering the correct key for SNOW-V using three attack methods. The y-axis represents the number of traces, while the x-axis demotes the method: CPA assisted with LDA, purely LDA-based analysis, and FCN with PCA based feature reduction. Results show that CPA+LDA requires 50 traces, LDA requires 12 traces, and FCN with PCA requires only 8 traces, demonstrating the superior efficiency of the FCN-based approach. }
    \label{fig:bar-Graph}
    \vspace{-2mm}
\end{figure}

To further contextualize our results, we present a bar graph comparing the three approaches—CPA+LDA, LDA-only, and FCN—as shown in \cref{fig:bar-Graph}. Recovering 8 bits of the secret key requires 50 traces with CPA+LDA, 12 traces with LDA, and only 8 traces with FCN combined with PCA, highlighting the superior efficiency of the FCN-based approach. Although FCN entails the highest training complexity, it achieves the best performance across all evaluation metrics with the fewest traces. This trade-off underscores the growing relevance of deep learning in side-channel analysis: the increased computational cost during training is offset by significant gains in trace efficiency and accuracy.

While this study demonstrates a machine learning-based power side-channel attack on the SNOW-V cipher implemented on an STM32 microcontroller using the ChipWhisperer platform, the methodology is broadly applicable to other platforms, devices, and cryptographic algorithms. The core principle-profiling power traces with machine learning models such as LDA and FCN-is independent of the specific hardware or algorithm. By identifying key-dependent leakage points in any target implementation and making minor adjustments to model parameters and preprocessing, this attack approach can be effectively adapted to a wide range of cryptographic algorithms.

%%%%%%%%%%%%%%%%%%%%%%%%%%%%%%%%%%%%%%%%%%%%%%%%%%%%%%%%%%%%%%

%%%%%%%%%%%%%%%%%%%%%%%%%%%%%%%%%%%%%%%%%%%%%%%%%%%%%%%%%%%%%%

\section{Discussions \& Conclusion}

In this paper, we presented a fully machine learning-based power SCA attack on the SNOW-V stream cipher. A TVLA was first conducted to confirm the presence of data-dependent leakage in the SNOW-V implementation on an STM32 microcontroller.

Compared to the previous work \cite{10545384}, which primarily relied on CPA combined with LDA, the current study introduces a more robust, model-agnostic methodology. While CPA+LDA (prior work) required a significantly higher number of traces ($\sim 50$) and was sensitive to power model accuracy, our proposed FCN+PCA-based approach eliminates this dependency with lower MTD ($\sim 8$), offering improved generalization in practical attack scenarios and demonstrating this using measured data on the SNOW-V running on the 32-bit ARM microcontroller.

Overall, our findings not only validate the vulnerability of SNOW-V to advanced profiling attacks but also highlight the need for robust countermeasures such as masking. Future work could investigate advanced neural architectures, automated feature selection techniques, and more extensive resilience testing across diverse leakage scenarios.

\bibliographystyle{unsrt}
\bibliography{Referance} % (make sure the filename is exactly the same)

\end{document}